\documentclass[a4paper]{article}
\usepackage{graphicx}
\usepackage{amsmath}
\usepackage{hyperref}
\usepackage{color}

\begin{document}

\begin{center}

{\bf \Large The norm game on a model network: a critical line}\\[5mm]

{\large  M. Rybak, A. Dydejczyk and K. Ku{\l}akowski }\\[3mm]

{\em

Faculty of Physics and Applied Computer Science,

AGH University of Science and Technology,

al. Mickiewicza 30, PL-30059 Krak\'ow, Poland

}

\bigskip

$^*${\tt fisher@autocom.pl, dydejczyk@ftj.agh.edu.pl, kulakowski@novell.ftj.agh.edu.pl}

\bigskip

\today

\end{center}

\begin{abstract}
The norm game (NG) introduced by Robert Axelrod is a convenient frame to disccuss the time evolution of the level 
of preserving norms in social systems. Recently NG was formulated in terms of a social contagion on a model social 
network with two stable states: defectors or punishers. Here we calculate the critical line between these states 
on the plane of parameters, which measure the severities of punishing and of being punished. We show also that the 
position of this line is more susceptible to the amount of agents who always punish and never defect, than to those 
who always defect and never punish. The process is discussed in the context of the statistical data on crimes in 
some European countries close to Wroc{\l}aw - the place of this Conference - around 1990.

\end{abstract}


\section{Introduction}

The physical boundary between a human being and his or her environment is the skin, but in the space of behaviours
the same boundary is less strict. Taking decisions, we are not completely selfish; we are to some extent bound by the social 
norms. The way of enforcing norms varies between a direct control and a deep internalization; in the latter case we treat
our conformity to norms in the same way as our payoff. Norms create the society, where we are formed \cite{mi}, and norms are 
modified by the society members. As it was formulated by a leading Polish psychiatrist Antoni K{\c e}pi\'nski, to decide where 
to put limits of our own rebel is one of most difficult problem in human life \cite{akp}. Solving this problem 
in our individual scale emerges in the social scale as a time evolution of norms. To search for laws 
which rule this process is a worthwhile challenge for the agent-based simulations. \\

A serious advance in this path was done by Robert Axelrod who formulated the norm game: an algorithm to simulate the 
conditions of persistence and fall of a social norm \cite{axel1,axel2}. The simulations done by Axelrod have been 
questioned \cite{galan}, but his paper has been cited hundreds times and it triggered a cascade of research;
for a recent review of simulations of norms see \cite{rev1}. Further, the subject of norms overlaps with the theory 
of cooperation; to cooperate is an example of a social norm. An overview on the latter might provide insight into 
current trends; still the research in this field seems to be at its intensively rising stage \cite{ax3,deff}. As norms 
are beliefs, there is also some overlap with the simulations of opinion dynamics; for a review of sociophysical simulations 
on this matter see \cite{san}. \\

Direct motivation of this research comes from statistical data on norm breaking. Perhaps most striking change we have seen
deals with the data on divorces in Portugal during the Carnival Revolution. There, the number of 
divorces increased from 777 in 1974 to 7773 in 1977 \cite{eur}. The plots on crime in countries in Central Europe are more
conventional. In Fig. 1 we show the data on Germany, Poland, Czech Republic, Austria and Hungary \cite{eur}.  In accordance 
with warnings by Eurostat, our aim is not to compare the amount of crimes in these countries, but rather to show the changes
in some of them. Note that opinion shifts 
were classified into continuous and abrupt by Michard and Bouchaud in 2005 \cite{bou} within a theory of imitation.
In the Axelrod model, the driving social mechanism is punishment; the interaction inhibits the change rather
than releases it.   \\

\begin{figure}[ht] 
\centering
{\centering \resizebox*{12cm}{8cm}{\rotatebox{-00}{\includegraphics{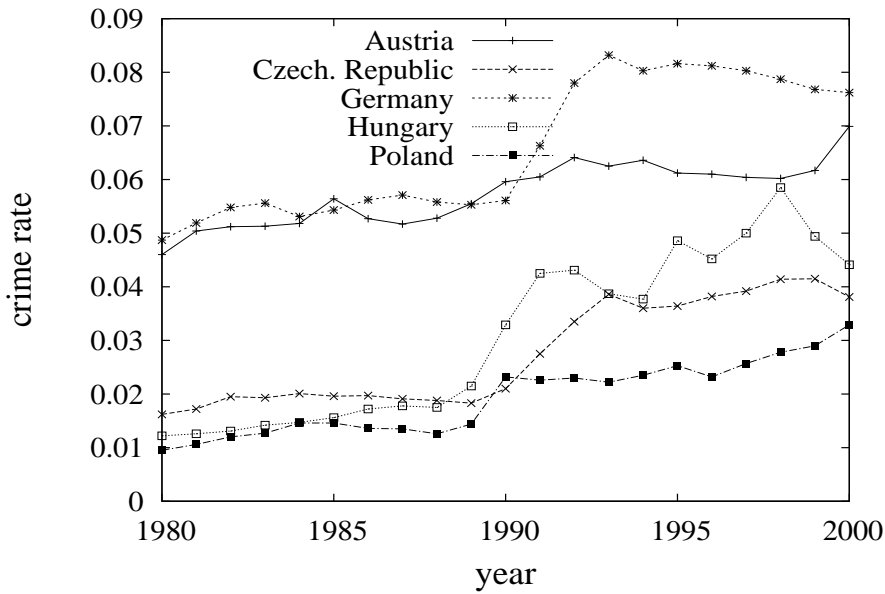}}}} 
\caption{The crime statistics in selected European countries in 1980-2000.}
\label{fig-1}
\end{figure}

\begin{figure}[ht] 
\centering
{\centering \resizebox*{12cm}{8cm}{\rotatebox{-00}{\includegraphics{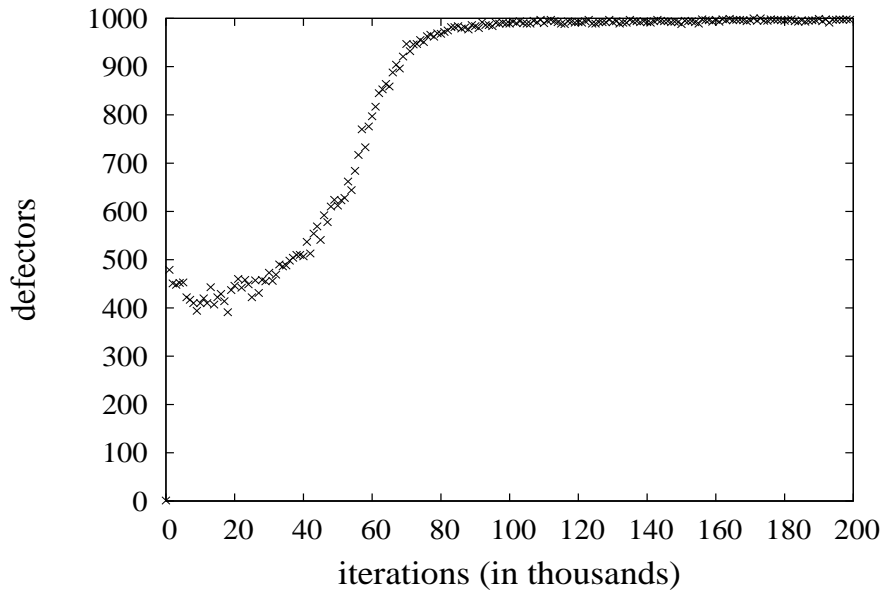}}}} 
\caption{Time evolution of the number of defectors - an example.}
\label{fig-2}
\end{figure}

\begin{figure}[ht] 
\centering
{\fbox{\centering \resizebox*{12cm}{8cm}{\rotatebox{-00}{\includegraphics{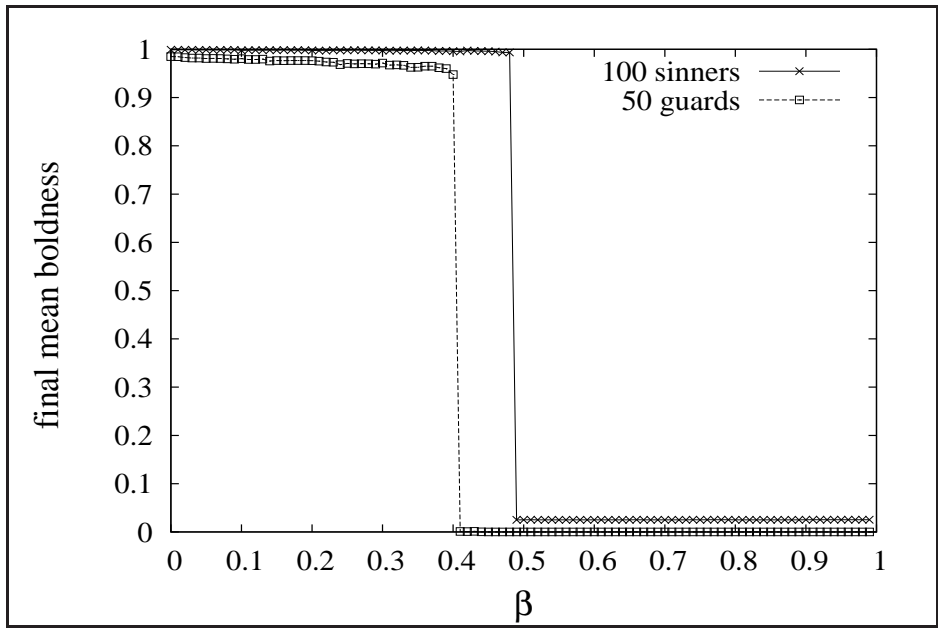}}}} }
\caption{The sharp character of the transition for a small number of guards or sinners.}
\label{fig-3}
\end{figure}

\begin{figure}[ht] 
\centering
{\centering \resizebox*{12cm}{8cm}{\rotatebox{-00}{\includegraphics{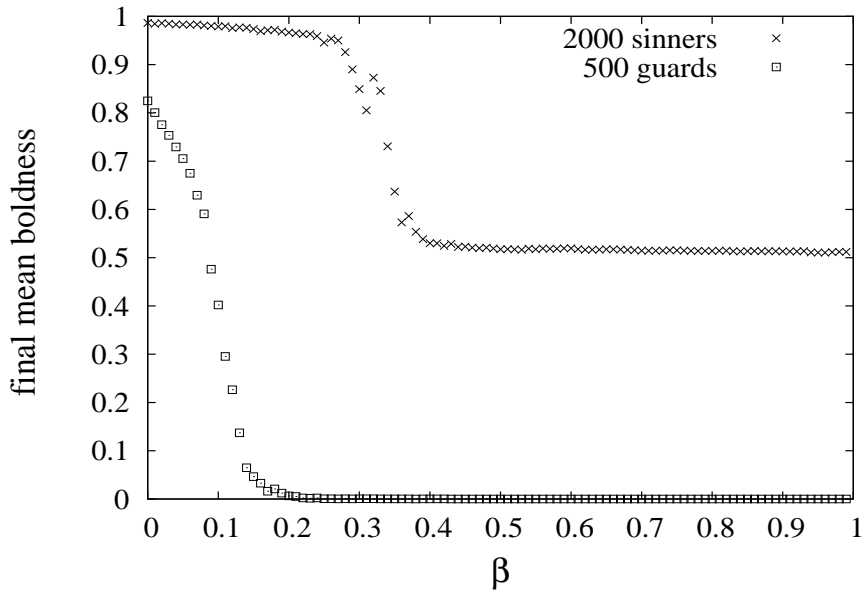}}}} 
\caption{The continuous character of the transition for a large number of guards or sinners.}
\label{fig-4}
\end{figure}

\begin{figure}[ht] 
\centering
{\centering \resizebox*{12cm}{8cm}{\rotatebox{-00}{\includegraphics{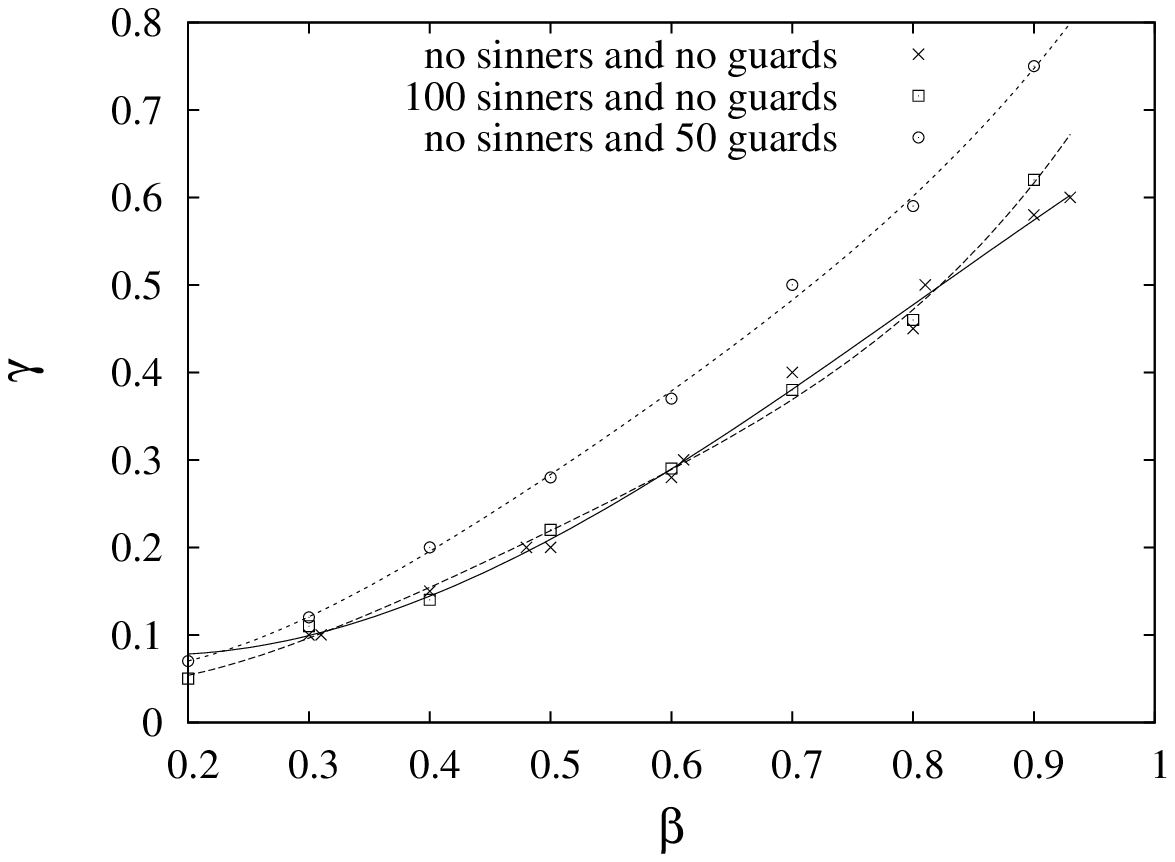}}}} 
\caption{The critical line in the homogeneous system and two lines for a small number of guards or sinners.}
\label{fig-5}
\end{figure}

In NG \cite{axel1} agents defect a norm with a given probability. Once an agent defects, other agents punish the defector, 
also with some probability. The defection is gratified with some payoff, but those who are punished lose. Also, those
who punish incur some cost. Axelrod considered also a metagame: the possibility of punishment those who do not punish.
The overall success of an agent was measured by his income, but the agent himself - represented by his strategy - was 
not modified. After some number of games, the genetic algoritm was used to select strategies which yielded the best 
income. As indicated in \cite{rev1}, this kind of modeling has an advantage to deal with the dynamics of the process.
We should add that Neumann criticizes the approach of Axelrod for disregarding the functional character of norms \cite{rev1}.\\

Recently \cite{kkk,kdr} we developed a new realization of the Axelrod model with two new ingredients. First is that the model 
is freed from the payoff parameters; what is left is just the influence of agents' decisions on decisions of other agents.
Further, the probabilities of decisions of individual agents (to defect or not, to punish or not) are not constant, but 
they are dynamically modified in each game they play. Second modification is less technical: once an agent decides to defect,
his ability to punish in his future games vanishes, and his probability to defect (boldness) in future games is kept one 
until he is punished; then it is multiplied by $(1-\beta)$. Then, the constant $\beta$ describes the severity of the punishment. 
On the contrary, once an agent punishes, he will never defect the norm, and his probability to punish (vengeance) is set to 
$(1-\gamma)$, where $\gamma$ is due to the punishment cost. This vengeance can be further reduced 
if the agent punishes also in his future games. In this way, a kind of social labeling takes place: first decision is 
irreversible, and the whole process can be seen as a social contagion \cite{san}. As a consequence, a sharp transition of 
the final boldness as dependent of the initial boldness is found \cite{kdr}. The threshold value varies with the model
parameters $(\beta,\gamma)$. These results depend only quantitatively on the assumed topology of the social network, which
determines the probability distribution of the number of punishers, i.e. of the node degree. \\

The aim of the present paper is twofold. First, we are going to investigate the above mentioned sharp transition
in the space of parameters $(\beta,\gamma)$. On the contrary to \cite{kdr}, here we are going to assume a given probability
distribution of the initial boldness between agents. This modification makes the calculation closer to a social reality,
where different agents present different willingnesses to break the norm and to punish. We should precise that in our model
the parameters $(\beta,\gamma)$ represent not the boldness and the vengeance, but rather the modifications of agents'
boldness and vengeance due to the decisions of other agents. Then, our model parameters describe the interactions between 
agents and not their actual states. Our goal here is to calculate the critical line on the plane $(\beta,\gamma)$ between
the final state 'all defect and nobody is punished' and  the final state 'nobody defects'. This critical line is a kind of
generalization of the critical concentration, calculated in the problems of directed percolation \cite{stauf,book}.
Our second aim is to investigate the character of the transition when some biased modifications of the structure of the 
social network. We use two kinds of these modifications: {\it i)} agents (guards) at some amount of nodes always punish 
and never defect, {\it ii)} agents (sinners) at some amount of nodes always defects and never punish.\\

In our model, agents are placed at nodes of the directed Erd\"os-Renyi network. This network is selected for its generic topology. 
There is 
much work on the structures of social networks, mostly done by sociologists \cite{cath,scott}; for an early list of 
references see \cite{hanrid}. With the outburst of scale-free networks a common opinion appeared that social networks are 
scale-free. Some of them can indeed be classified as scale-free, in particular those where direct face-to-face contact 
is not needed; as citation networks or telephone-calls networks \cite{mark1,kert1,kert2}. Still, the actual structure varies 
from one social network to another \cite{mark2,ama}, and often the network is simply too small to be classified to any type 
\cite{oddie}. To end, we have checked \cite{kdr} that the investigated threshold appears also in the scale-free growing 
networks, except the case when the direction of links is determined by the sequence of attaching new nodes. Last but not least, 
here we are not going to discuss the role of hubs, which could complicate the results.\\

The paper is organized as follows. In the next section the model assumptions and the details of the calculations are listed. 
In Section 3 we describe the numerical results. These are two: the critical line on the plane of the parameters, and the 
transition dependence of the amount of guards/sinners. Section 4 is devoted to discussion of the results in the context 
of a recent classification of contagion processes \cite{dowa}, and of some statistical data on dynamics of crime, presented 
in Fig. 2.

\section{Model and calculations}

The network size is $N=4000$ nodes, the mean number of in-going links (punishers) is $\lambda=5$, and their distribution is 
Poissonian. On the contrary to our former calculations \cite{kdr}, here we assume a homogeneous distribution of the initial 
probability to defect the norm, i.e. the initial boldness $b(i)$, where $i$ is the node index. As a rule, the initial 
vengeance $v(i)$, i.e. the initial probability of punishing, is $v(i)=1-b(i)$. This condition is not maintained during
the simulation; however, $b(i)+v(i)\le 1$. Third option is to obey the norm and not punish, with the probability $1-b(i)-v(i)$. \\

At each time step, an agent $i$ is selected and he breaks the norm with the probability equal to his boldness $b(i)$. 
If actually he does, his boldness $b(i)$ is set to 1 and his vengeance, i.e. the probability of punishing - is set to 0.
Then his neighbours $j(i)$ are asked, one by one, if they punish $i$. If one of them punishes, the boldness of $i$ is 
multiplied by a factor $1-\beta$, and the vengeance of the punisher $j$ is multiplied by $1-\gamma$. The defector can be 
punished only by one neighbour. On the contrary, if a neighbour does not punish, his boldness is set to one and his vengeance 
is set to zero. In this way, the process is accompanied by  kind of social labeling \cite{oxf}: those who break the norm and 
those who refrain from punishing cannot punish in their future games, and those who punish cannot break the norm.\\

As a rule, we calculate the values of the parameters $\beta, \gamma$ where the threshold appears, i.e. the final state changes 
from the bold state 'all defect' to the vengeant state 'all punish'. Additionally, as a new variant of the game, some amount 
of sites of the network is selected randomly. Agents at these nodes got special 
roles of 'guards' or 'sinners'. If an agent is a guard, he always punishes and never defects; sinners do the opposite, i.e.
always defect and never punish. The value and character of the threshold is observed against the ratio of the number of 
those special nodes to the whole population $N$. All these special nodes are either all 'guards' or all 'sinners'.

\section{Results} 

Calculating the final boldness as dependent on $\beta$ and $\gamma$ we observe a sharp change of the result, as in Fig. 3 a,
at some threshold values of $\gamma$ and $\beta$. This means that we got a critical line $\gamma_c(\beta)$ in the plane 
of the parameters $(\beta,\gamma)$ - see Fig. 4. This line divides the plane $(\beta,\gamma)$ into two areas; the plane can 
be treated as a phase diagram, then we can talk about two phases. Abobe the line, we have a 'Bold' phase, where $\gamma$ is 
large; there punishment costs too much and is not effective. Below the line, we have a 'Vengeant' phase where the cost of
punishment is low. Then, everybody punishes and there is no interest in defection. \\

Having added a small amount of 'sinners' or 'guards' to the system we observe that the threshold changes differently 
in these two cases. Basically the character of the threshold remains the same, just the threshold value is more susceptible
to the admixture of 'guards' than to the one of 'sinners'. This can be seen at the positions of two additional curves in 
Fig. 4. While adding of five percent of 'sinners' apparently produces no effect, twice smaller admixture of 'guards' shifts
the critical line upwards, reducing the area of the Bold phase. When the number of modified nodes increases more, the 
character of the plot $b(\gamma;\beta)$ gradually changes from an abrupt to a more continuous one. Examples of these curves 
are shown in Fig. 3 b. This change prevents us to investigate the critical line for higher concentration of special nodes; 
the transition becomes fuzzy.

\section{Discussion}

The crossover from the sharp to the fuzzy character of the transition between the Bold phase and the Vengeant phase, observed at the 
$(\beta,\gamma)$ plane, fits into a recent classification of contagion processes \cite{dowa}. According to that scheme, models of 
contagion processes can be divided into three classes: $I)$ independent interaction models, $II)$ stochastic threshold models,
and $III)$ deterministic threshold models. Sharp transitions, like those found in our results, are characteristic for class $III$.
Models in class $II$ give fuzzy curves, similar to those obtained here for larger amounts of admixtured 'guards' or 'sinners'.
These models are also called 'critical mass models'. In our terms, a critical amount of sinners should not be punished to get the 
transition to the bold phase; this could mean that local concentration of 'guards' should be small. If this is so, our results 
do fit into the classification proposed in \cite{dowa}. As we noted in the Introduction, a similar classification was developed
in \cite{bou}. Note that the basis of \cite{bou} was the random-field Ising model, which is far from the picture of contagion. \\

Coming back to the sociological reality, let us add a few words on the data presented in Fig. 1 in terms of punishment and its cost.
We stress that the parameters $\beta$ and $\gamma$ do not mean directly the amount of units which a punished and punishing agent 
should pay, but just the measure of the decrease of the probability that he will defect and punish again. In these terms, large 
numbers on the statistics of crimes in a country can be interpreted as an indication that the punishment cost are large or the 
punishment is weak. Accordingly, an increase of the data can mean that the punishment $\beta$ decreased below some critical value 
or the punishment cost $\gamma$ increased. Then, new generations faced with an issue, to break a given norm or to preserve it, 
decide in a collective way, and these decisions are visible in the statistical data. If this point of view is accepted, we should 
admit in particular that law in Germany, Hungary and Poland is broken much more frequently after 1993. However, the data presented 
in Fig. 1 can be seen also as a demonstration, that police in these countries is less punished for detecting crimes after 1993, than 
before. Obviously, this punishment is not open and intended; still it can appear as a consequence of burdensome bureaucratic procedures, 
faulty organization, unclear rights and shifts of political aims.

\bigskip

{\bf Acknowledgements.} The research is partially supported within the FP7 project SOCIONICAL, No. 231288.

\end{document}